\RequirePackage{fix-cm}
\documentclass[smallextended]{svjour3}       
\smartqed  
\usepackage{graphicx}
\usepackage[ansinew]{inputenc}
\usepackage{amssymb}
\usepackage{amsmath}

\begin{document}

\def\R{\mathbb{R}}
\def\N{\mathbb{N}}
\def\Z{\mathbb{Z}}
\def\Q{\mathbb{Q}}
\def\C{\mathbb{C}}

\newtheorem{thm}{Theorem}
\newtheorem{lem}{Lemma}
\newtheorem{cor}{Corollary}
\newtheorem{prop}{Proposition}
\newtheorem{defin}{Definition}
\newtheorem{rmk}{Remark}
\newtheorem{prob}{Problem}
\newtheorem{con}{Conjecture}
\newtheorem{ex}{Example}

\def\sobre#1#2{\smash{\mathop{#1}\limits^{#2}}} 
\def\bajo#1#2{\smash{\mathop{#1}\limits_{#2}}}  

\font\errefont=cmssi10 scaled 920
\def\C{\mathop{\,\mbox{\errefont I}\!\!C}\nolimits}

\title{Quantum codes do not fix isotropic errors}

\subtitle{}

\author{J. Lacalle         \and
        L.M. Pozo-Coronado \and
        A.L. Fonseca de Oliveira 
}

\institute{J. Lacalle \at
              Dep. de Matem\'atica Aplicada a las TIC, ETS de Ingenier\'{\i}a de Sistemas Inform\'aticos, Universidad Polit\'ecnica de Madrid, C/ Alan Turing s/n, 28031, Madrid, Spain \\
              \email{jesus.glopezdelacalle@upm.es}
           \and
           L.M. Pozo-Coronado \at
              Dep. de Matem\'atica Aplicada a las TIC, ETS de Ingenier\'{\i}a de Sistemas Inform\'aticos, Universidad Polit\'ecnica de Madrid, C/ Alan Turing s/n, 28031, Madrid, Spain \\
              \email{lm.pozo@upm.es}
           \and
           A.L. Fonseca de Oliveira \at
              Facultad de Ingenier\'{\i}a, Universidad ORT Uruguay, Montevideo, Uruguay \\
              \email{fonseca@ort.edu.uy}
}

\date{Received: date / Accepted: date}

\maketitle

\begin{abstract}
In this work we prove that quantum error correcting codes do not fix isotropic errors (Theorem~\ref{Thm:FinalResult}), even assuming that their correction circuits do not introduce new errors. We say that a quantum code does not fix a quantum computing error if its application does not reduce the variance of the error.
We also prove for isotropic errors that, if the correction circuit of a quantum code detects an error, the corrected logical $m-$qubit has uniform distribution (Theorem~\ref{Thm:Isotropy}) and as a result, it already loses all the computing information.

\end{abstract}

\section{Introduction}

It is well known that the main challenge to achieve an efficient quantum computation is the control of quantum errors~\cite{Ga}. To address this problem, two fundamental tools have been developed: quantum error correction codes~\cite{CS,St1,Go1,CRSS} in combination with fault tolerant quantum computing~\cite{Sh,Pr1,St2,Go2,GHW,Na}.

In this article we study the effectiveness of an arbitrary quantum error correcting code to fix isotropic errors in quantum computing. In order to do that, we represent $n-$qubits as points of the unit real sphere of dimension $d=2^{n+1}-1$~\cite{NC}, $S^d=\{x\in\R^{d+1}\ |\ \|x\| =1\}$, taking coordinates with respect to the computational basis $[|0\rangle,|1\rangle,\dots,|2^n-1\rangle]$,
\begin{eqnarray}\label{QubitFormula}
\Psi=(x_0+ix_1,x_2+ix_3,\dots,x_{d-1}+ix_{d}).
\end{eqnarray}

Following a previous work~\cite{LP}, we consider quantum computing errors as random variables with density function defined on $S^d$. As mentioned in that article, it is easy to relate this representation to the usual representation in quantum computing by density matrices. In fact, if $X$ is a quantum computing error with density function $f(x)$, then the density matrix of $X$, $\rho(X)$, is obtained as follows, using the pure quantum states given by Formula (\ref{QubitFormula}):
$$
\rho(X)=\int_{S^d}f(x)|\Psi\rangle\langle\Psi|dx\quad\text{where}\quad\int_{S^d}f(x)dx=1.
$$

Density matrices do not always discriminate different quantum computing errors~\cite{NC}. Therefore, representations of quantum computing errors by random variables are more accurate than those by density matrices. Beside other considerations, while the space of random variables over $S^d$ is infinite-dimensional, the space of $n-$qubit density matrices has finite dimension. This is the main reason why the authors decided to use random variables to represent quantum computing errors. And once the representation of quantum computing errors is established by random variables, the most natural parameter to measure the size of quantum computing errors is the variance.

As described in~\cite{LP}, the variance of a random variable $X$ is defined as the mean of the quadratic deviation from the mean value $\mu$ of $X$, $V(X)=E[\|X-\mu\|^2]$. In our case, since the random variable $X$ represents a quantum computing error, the mean value of $X$ is the $n-$qubit $\Phi$ resulting from an errorless computation. Without loss of generality, we will assume that the mean value of every quantum computing error will always be $\Phi=|0\rangle$. To achieve this, it suffices to move $\Phi$ into $|0\rangle$ through a unitary transformation. Therefore, using the pure quantum states given by Formula (\ref{QubitFormula}), the variance of $X$ will be
$$
V(X)=E[\|\Psi-\Phi\|^2]=E[2-2x_0]=2-2\int_{S^d}x_0f(x)dx.
$$

In~\cite{LP} the variance of the sum of two independent errors on $S^d$ is presented for the first time. It is proved for isotropic errors and it is conjectured in general that
\begin{eqnarray}\label{VarFormula}
V(X_1+X_2)=V(X_1)+V(X_2)-\frac{V(X_1)V(X_2)}{2}.
\end{eqnarray}

To relate the variance to the most common error measure in quantum computing, fidelity~\cite{NC}, the authors define a quantum variance that takes into account that quantum states are equivalent under multiplication by a phase. Thereby, the quantum variance of a random variable $X$ is defined as:
$$
V_q(X)=E[\bajo{\text{min}}{\phi}(\|\Psi-e^{i\phi}\Phi\|^2)]=2-2E\left[\sqrt{x_0^2+x_1^2}\right].
$$

The fidelity of the random variable $X$, $F(X)$, with respect to the pure quantum state $\Phi=|0\rangle$ satisfies $F(X)=\sqrt{\langle \Phi|\rho(X)|\Phi\rangle}$~\cite{NC}. Therefore, using the pure quantum states given by Formula (\ref{QubitFormula}), $F(X)^2=E[\langle \Phi|\Psi\rangle\langle \Psi|\Phi\rangle]=E[|\langle \Phi|\Psi\rangle|^2]=E[x_0^2+x_1^2]$. Now, the property $\sqrt{x_0^2+x_1^2}\geq x_0^2+x_1^2$ and Jensen's inequality $\sqrt{E[x_0^2+x_1^2]}\geq E\left[\sqrt{x_0^2+x_1^2}\right]$ allow us to conclude that:
$$
1-\dfrac{V_q(X)}{2}\leq F(X)\leq\sqrt{1-\dfrac{V_q(X)}{2}}.
$$

These inequalities show that quantum variance and fidelity are essentially equivalent, since when quantum variance tends to $0$, fidelity tends to $1$ and, conversely, when fidelity tends to $1$, quantum variance tends to $0$. Of the three measures, the one that best reflects the distribution of the $n-$qubit error is the variance. Hence, the goal of this work is to study the behavior of the variance, as a measure of error, when using quantum error correcting codes.

The study of the effectiveness of quantum error correcting codes has become essential to face the challenge of quantum computing. In this context, the problem we want to address is the following: Let $\Phi$ be an $n-$qubit encoded by a quantum code $\cal C$. Suppose that the coded state $\Phi$ is changed by error, becoming the state $\Psi$. Now, to fix the error we apply the code correction circuit, obtaining the final state $\tilde\Phi$. While $\Phi$ is a pure state, $\Psi$ and $\tilde\Phi$ are random variables (mixed states). Our goal is to compare the variance of $\tilde\Phi$, $V(\tilde\Phi)=E\left[\|\tilde\Phi-\Phi\|^2\right]$, with that of $\Psi$, $V(\Psi)=E\left[\|\Psi-\Phi\|^2\right]$.

In order to compare the variances we will assume that the corrector circuit of $\cal C$ does not introduce new errors. In other words, we are going to estimate the theoretical capacity of the code to correct quantum computing errors. One would ideally expect that $\tilde\Phi=\Phi$ so that the variance of $\tilde\Phi$ would be $V(\tilde\Phi)=0$. Being more practical, we are only going to demand the minimum that could possibly be asked from an error correction process: $V(\tilde\Phi)<V(\Psi)$. If this minimum requirement is not met, we will say that the code $\cal C$ does not fix the corresponding quantum computing error.

The problem we address is, in our opinion, one of the biggest challenges for quantum computing. Consequently, it is also one of the most difficult tasks. For this reason, we restrict the problem to a certain type of error that allows us to effectively compute the variances of the disturbed and corrected states, $\Psi$ and $\tilde\Phi$. Specifically, we study the isotropic quantum computing errors introduced in~\cite{LP}. They are errors whose density function (as random variables) depends exclusively on $\|\Phi-\Psi\|$. We shall prove that no code fixes isotropic quantum computing errors (Theorem~\ref{Thm:FinalResult}) and that, for this type of errors, if the correction circuit of a quantum code detects an error, the corrected logical $m-$qubit has uniform distribution (Theorem~\ref{Thm:Isotropy}) and as a result, it already loses all the computing information.

There are many works related to the control of quantum computing errors, in addition to those already mentioned above. General studies and surveys on the subject~\cite{Sc,Pr2,Go3,CDT,HFWH,DP,HZKBR,HKR}, about the quantum computation threshold theorem~\cite{AGP,WFSH,ACGW,CT}, quantum error correction codes~\cite{OG,LZLX,CCHF,Gu}, concatenated quantum error correction codes~\cite{BPFHC,ES} and articles related to topological quantum codes~\cite{DPB,NJS}. Lately, quantum computing error control has focused on both coherent errors~\cite{GM,BEKP} and cross-talk errors~\cite{PSVW,BTS}. Finally, we cannot forget the hardest error to control in quantum computing, the quantum decoherence~\cite{Zu}. As we have commented above, these quantum computing errors can be analyzed in the framework of random variables that has been set in~\cite{LP}. In the conclusions we analyze in more detail the characteristics of the different types of error from the point of view of their control and in view of the main result obtained in this paper.

The outline of the article is as follows: in Section 2 we set up the general structure of quantum error correcting codes; in Section 3 we analyze isotropic quantum computing errors, we calculate the variance of the disturbed state $\Psi$ and we introduce the normal distribution, as a particular case of isotropic distribution; in Section 4 we establish the result of applying the correction circuit of a quantum error correcting code to an isotropic quantum computing error, we prove that no code fixes isotropic quantum computing errors and we analyze the behavior of an isotropic error with normal distribution; finally, in Section 5 we analyze the characteristics of the different types of quantum computing error, relating them to the main result obtained in this article.

\section{Quantum error correcting codes}

An error-correcting quantum code of dimension $[n, m]$ is a subspace $\cal C$ of dimension $d^{\prime}=2^m$ in the $n-$qubit space ${\cal H}^n$, whose dimension is $d=2^n$. The $\cal C$ quantum code encoding function is a unitary operator $C$ that satisfies the following properties:
$$
C:\,{\cal H}^m\otimes{\cal H}^{n-m}\,\to\,{\cal H}^n \ \text{and}\ {\cal C}=C({\cal H}^m\otimes|0\rangle).
$$

The $\cal C$ code fixes $d^{\prime\prime}=2^{n-m}$ discrete errors: $E_0,\ E_1,\ \dots,\ E_{d''-1}$. Since the identity $I$ should be among these unitary operators, we assume that $E_0=I$. This process of discretization of errors allows to correct any of them if the subspaces $S_s=E_s({\cal C})$, $0\leq s < d^{\prime\prime}$, satisfy the following property:
\begin{equation}
\label{For:OrthogonalSum}
{\cal H}^n=S_0\,\bot\,S_1\,\cdots\,\bot\,S_{d''-1}.
\end{equation}
That is, ${\cal H}^n$ is the orthogonal direct sum of said subspaces. Note also that $S_0=E_0({\cal C})=I({\cal C})={\cal C}$. In the stabilized code formalism, the code $\cal C$ is the subspace of fixed states of an abelian subgroup of the Pauli group ${\cal P}_n=\{\pm 1,\pm i\}\times\{I,X,Z,Y\}^n$ and discrete errors are operators of ${\cal P}_n$ that anti-commute with any of the subgroup generators, except for the identity operator $E_0$. If Formula~(\ref{For:OrthogonalSum}) holds, the code is non-degenerate.

Suppose that a coded state $\Phi$ is changed by error, becoming the state $\Psi$. The initial state is a code state, that is, $\Phi\in S_0$, while the final state in general is not, that is, $\Psi\not\in S_0$. If the disturbed state belongs to the subspace $W_{\Phi}=L(E_0\Phi,\dots,E_{d''-1}\Phi)$, that is, if it is of the form
\begin{equation}
\label{For:FixableErrors}
\Psi=\alpha_0E_0\Phi+\cdots+\alpha_{d''-1}E_{d''-1}\Phi\quad\text{with}\quad|\alpha_0|^2+\cdots+|\alpha_{d''-1}|^2=1,
\end{equation}
then the quantum code allows us to retrieve the initial state $\Phi$. To achieve this, we measure $\Psi$ with respect to the orthogonal decomposition of the Formula~(\ref{For:OrthogonalSum}). The result will be $\frac{\alpha_s}{|\alpha_s|}E_s\Phi$ for a value $s$ between $0$ and $d^{\prime\prime}-1$. The value of $s$ is called syndrome and allows us to identify the discrete error that the quantum measurement indicates. Then, applying the quantum operator $E_s^{-1}$ we obtain $\frac{\alpha_s}{|\alpha_s|}\Phi$. This state is not exactly $\Phi$ but, differing only in a phase factor, both states are indistinguishable from the point of view of Quantum Mechanics. Therefore, the code has fixed the error.

An error that does not satisfy Formula~(\ref{For:FixableErrors}), that is, it does not belong to $W_{\Phi}$, cannot be fixed exactly. For example, if $\Psi$ belongs to the code subspace $\cal C$, the error cannot be fixed at all since, being a code state, it is assumed that it has not been disturbed. In this work we want to analyze the limitation in the correction capacity of an arbitrary code, assuming that the code correction circuit does not introduce new errors.

Finally, we want to highlight that discrete errors can be chosen so that, for example, all errors affecting a single qubit are fixed. The best code with this feature that encodes one qubit is a $5$-qubit code~\cite{BDSW,LMPZ}. This code is optimal in the sense that no code with less than $5$ qubits can fix all the errors of one qubit.

\section{Isotropic quantum computing errors}

As in the previous section, let $\Phi$ be an initial code state, let $\cal C$ be the code with which $\Phi$ has been encoded and let $\Psi$ be the final state caused by an error on $\Phi$. In the $n-$qubit space ${\cal H}^n$ we consider the basis ${\cal B}_n=\left[\,|0\rangle,\,|1\rangle,\,\dots\,|d-1\rangle\,\right]$ in such a way that the following properties hold:
\begin{itemize}
\item[a)] $\Phi=|0\rangle$ and $S_0=L(|0\rangle,\dots,|d^{\prime}-1\rangle)$.
\item[b)] $E_s\Phi=|sd^{\prime}\rangle$ for all $0\leq s < d^{\prime\prime}$.
\item[c)] $E_s|k\rangle=|sd^{\prime}+k\rangle$ for all $0\leq s < d^{\prime\prime}$ and $0\leq k < d^{\prime}$.
\item[d)] $S_s=L(|sd^{\prime}\rangle,|sd^{\prime}+1\rangle\dots,|sd^{\prime}+d^{\prime}-1\rangle)$ for all $0\leq s < d^{\prime\prime}$.
\end{itemize}

The state $\Psi$ can be represented, with respect to the basis $B_n$, as follows:
\begin{equation}
\label{For:BinomialForm}
\Psi=\sum_{k=0}^{d-1}\alpha_k|k\rangle\quad\text{such that}\quad \sum_{k=0}^{d-1}|\alpha_k|^2=1.
\end{equation}

If we represent the coefficients in binomial form, that is, $\alpha_k = x_{2k} + i\,x_{2k + 1}$ for all $0\leq k< d$, the state $\Psi$ is parameterized by a point of a unit sphere of dimension $2d-1$ that we denote by ${\cal S}_{2d-1}$:
$$
\Psi\ \equiv\ (x_0,\,x_1,\,\dots,\,x_{2d-1})\quad\text{such that}\quad \sum_{j=0}^{2d-1}x_j^2=1.
$$

We use this parametrization to introduce the probability distribution of the error. However, we don't use Cartesian coordinates but spherical coordinates:
$$
\Psi\ \equiv\ (\theta_0,\,\theta_1,\,\dots,\,\theta_{2d-2})\quad\text{such that}\quad
\left\{\begin{array}{l}
0\leq\theta_0,\,\dots,\,\theta_{2d-3}\leq\pi \\ \\
0\leq\theta_{2d-2}\leq2\pi
\end{array}\right.,
$$

\begin{equation}
\label{For:CartesianVsSpherical}
\begin{array}{l}
x_j=\sin(\theta_0)\,\cdots\,\sin(\theta_{j-1})\,\cos(\theta_j)\quad\text{for all}\quad 0\leq j\leq 2d-2, \\ \\
x_{2d-1}=\sin(\theta_0)\,\cdots\,\sin(\theta_{2d-2}).
\end{array}
\end{equation}

Let $\tilde{\Phi}$ be the code state resulting from applying the code correction circuit to the state $\Psi$, assuming that this circuit does not introduce new errors. From the point of view of the statistical study of errors, the disturbed state $\Psi$ is a random variable. The same holds for the state $\tilde{\Phi}$ resulting from the correction. The random variable $\Psi$ describes the distribution of the error around $\Phi$ in the sphere ${\cal S}_{2d-1}$. On the other hand, the random variable $\tilde{\Phi}$ describes the distribution of the error after correction around $\Phi$ in the sphere ${\cal S}_{2d^\prime -1}$ of the code subspace $\cal C$ (since the accuracy of the correction circuit we are assuming implies that $\tilde{\Phi}$ belongs to $\cal C$).

In this context, in order to measure the code correction capacity, we compare the variance of the final error $\tilde{\Phi}$, with the variance of the initial error $\Psi$. Our study focuses on isotropic errors so, to describe the problem precisely, we introduce the definitions of isotropic errors and their variances.

\begin{defin}
The random variable $\Psi$ is isotropic if its density function depends exclusively on $\|\Phi-\Psi\|$.
\end{defin}

Considering that $\Phi=|0\rangle$ and Formula~(\ref{For:BinomialForm}) with the coefficients in binomial form, it follows that:
$$
\begin{array}{ccl}
\|\Phi-\Psi\|^2 & = &|(x_0-1)+ix_1|^2+|x_2+ix_3|^2+\cdots +|x_{2d-2}+ix_{2d-1}|^2 \\
                  & = & \displaystyle 1-2x_0+\sum_{k=0}^{2d-1}x_j^2=2-2x_0. \\
\end{array}
$$

Finally, taking into account Formula~(\ref{For:CartesianVsSpherical}), it is concluded that:
$$
\|\Phi-\Psi\|^2 = 2-2\cos(\theta_0).
$$

\begin{defin}
The variance of the isotropic random variable $\Psi$ is the expected value:
$$
E\left[\|\Phi-\Psi\|^2\right] = 2-2E\left[\cos(\theta_0)\right].
$$
\end{defin}

\begin{thm}
The variance of an isotropic random variable $\Psi$ with density function $f(\theta_0)$ is equal to:
$$
V(\Psi) = 2 - 4 \dfrac{(2\pi)^{d-1}}{(2d-3)!!}{\bar E}\left[\cos(\theta_0)\sin^{2d-2}(\theta_0)\right],
$$
where $\displaystyle {\bar E}\left[\cos(\theta_0)\sin^{2d-2}(\theta_0)\right]=\int_0^\pi f(\theta_0)\cos(\theta_0)\sin^{2d-2}(\theta_0)d\theta_0$.
\label{Thm:Variance}
\end{thm}
\begin{proof}
The expected value is calculated by integrating over the sphere ${\cal S}_{2d-1}$. This integral, in spherical coordinates, is:
\begin{eqnarray*}
V(\Psi) & = & 2 - 2 \displaystyle\int_{{\cal S}_{2d-1}} f(\theta_0)\cos(\theta_0)d_{{\cal S}_{2d-1}} \\
& = & 2 - 2 \displaystyle |{\cal S}_{2d-2}| \int_0^\pi f(\theta_0)\cos(\theta_0)\sin^{2d-2}(\theta_0) d\theta_0, \\
\end{eqnarray*}
where $\sin^{2d-2}(\theta_0)$ is the volume element corresponding to $\theta_0$ and $|{\cal S}_{2d-2}|$ is the volume of the sphere of dimension $2d-2$.

Finally, the theorem is proved by substituting the value of $|{\cal S}_{2d-2}|$ (see Appendix) in the previous expression:
$$
|{\cal S}_{2d-2}| = 2\dfrac{(2\pi)^{d-1}}{(2d-3)!!}.
$$
\end{proof}

From now on ${\bar E}[g(\theta_0)]$ will denote the expected value of the function $g(\theta_0)$ on the interval $[0,\pi]$, for isotropic density functions $f(\theta_0)$. Since $f(\theta_0)$ is a density, then the following equality holds:
\begin{equation}
\label{For:DensityFunction}
{\bar E}[\sin^{2d-2}(\theta_0)] = \dfrac{1}{|{\cal S}_{2d-2}|} = \dfrac{(2d-3)!!}{2(2\pi)^{d-1}}.
\end{equation}

\subsection{An example}
\label{Sub:Example}

In this subsection we introduce a special distribution of isotropic error that has been key for~\cite{LP} and this article.

\begin{defin}
The normal error distribution is one that has the following density function:
$$
f_n(\sigma,\theta_0)=\frac{(2d-2)!!}{(2\pi)^d}\frac{(1-\sigma^2)}{(1+\sigma^2-2\sigma\cos(\theta_0))^d},
$$
where the parameter $\sigma$ belongs to the interval $[0,1)$.
\end{defin}

\begin{figure}[th]
\label{Fig:Distr}
\begin{center}
        \includegraphics[scale=1]{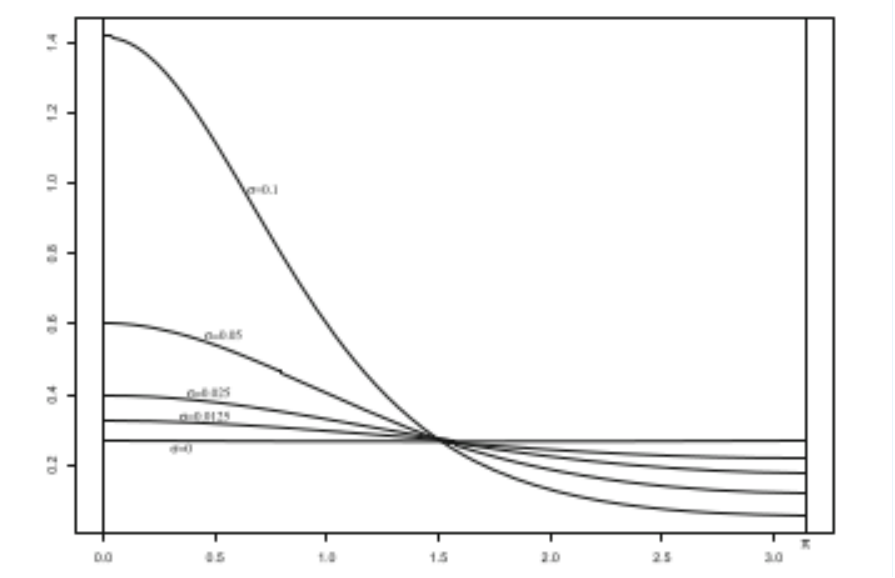}
        \caption{\centerline{Normal density function for $d=8$.}}
\end{center}
\end{figure}

When $\sigma$ approaches $1$ the probability is concentrated at the point $\Phi$, canceling the error. And when $\sigma$ approaches $0$ the distribution tends to be uniform, that is, a distribution in which, after the disturbance, all the states are equally probable. Figure~\ref{Fig:Distr} shows how the distribution changes depending on the parameter.

For us, the most important property of an error distribution is its variance. In the case of the normal distribution its variance has a very simple expression.

\begin{thm}
\label{Thm:VarianceNormal}
The variance of a normal error distribution $\Psi$ is:
$$
V(\Psi)=2(1-\sigma).
$$
\end{thm}

\begin{proof}
Applying Theorem~\ref{Thm:Variance} and substituting the integral for its value (see Appendix) we obtain:
\begin{eqnarray*}
V(\Psi) & = & \displaystyle 2 - 4 \dfrac{(2\pi)^{d-1}}{(2d-3)!!}\int_0^\pi f_n(\theta_0)\cos(\theta_0)\sin^{2d-2}(\theta_0)d\theta_0 \\ \\
 & = & \displaystyle 2-4\ \dfrac{(2\pi)^{d-1}}{(2d-3)!!}\dfrac{(2d-2)!!}{(2\pi)^d}(1-\sigma^2)\int_0^\pi \dfrac{\cos(\theta_0)\sin^{2d-2}(\theta_0)}{(1+\sigma^2-2\sigma\cos(\theta_0))^d}d\theta_0 \\ \\
 & = & \displaystyle 2-4\ \frac{(2d-2)!!\ (1-\sigma^2)}{2\pi\ (2d-3)!!}\ \frac{(2d-3)!!}{(2d-2)!!}\frac{\sigma}{(1-\sigma^2)}\pi \\ \\
 & = & 2-2\sigma.
\end{eqnarray*}
\end{proof}

\section{Result of applying the correction circuit of a quantum code to an isotropic quantum computing error}

When applying the correction circuit of code $\cal C$ to the disturbed state $\Psi$, the corrected state $\tilde{\Phi}$ is obtained in different ways, depending on the measured syndrome $s$ ($0\leq s< d^{\prime\prime}$). We denote by $P_s$ the probability that the syndrome is $s$, $0\leq s< d^{\prime\prime}$. Thus $P_0$ is the probability that no error is detected, that is, that the quantum measurement projects the disturbed state $\Psi$ on the subspace $S_0={\cal C}$. On the other hand, $P_s$ is the probability that the $E_s$ error will be detected for $0< s< d^{\prime\prime}$. In this case, after the measurement, the operator $E_s^{-1}$ must be applied. In order to calculate the probabilities $P_s$, $0\leq s< d^{\prime\prime}$, we need the following lemma.

\begin{lem}
\label{Lem:ExpectedCoordinates}
Given an isotropic random variable $\Psi$ with density function $f(\theta_0)$, the following statements hold true:
\begin{enumerate}
\item[a)] $\displaystyle E[x_0^2]=2\dfrac{(2\pi)^{d-1}}{(2d-3)!!}\ {\bar E}[\cos^2(\theta_0)\sin^{2d-2}(\theta_0)]$.
\item[b)] $\displaystyle E[x_j^2]=2\dfrac{(2\pi)^{d-1}}{(2d-1)!!}\ {\bar E}[\sin^{2d}(\theta_0)]$\quad for all\quad $0< j< 2d$.
\item[c)] $\displaystyle E[|\alpha_0|^2]=2\dfrac{(2\pi)^{d-1}}{(2d-3)!!}\left({\bar E}[\cos^2(\theta_0)\sin^{2d-2}(\theta_0)]+\dfrac{{\bar E}[\sin^{2d}(\theta_0)]}{2d-1}\right)$.
\item[d)] $\displaystyle E[|\alpha_k|^2]=4\dfrac{(2\pi)^{d-1}}{(2d-1)!!}\ {\bar E}[\sin^{2d}(\theta_0)]$\quad for all\quad $0< k< d$.
\end{enumerate}
\end{lem}
\begin{proof}
To prove item a), simply substitute in the following expression the value of $|{\cal S}_{2d-2}|$ given in the Appendix:
$$
\begin{array}{ccl}
E[x_0^2] & = & \displaystyle E[\cos^2(\theta_0)]=|{\cal S}_{2d-2}|\ \int_0^\pi f(\theta_0)\cos^2(\theta_0)\sin^{2d-2}(\theta_0)d\theta_0 \\
 & = & \displaystyle 2\dfrac{(2\pi)^{d-1}}{(2d-3)!!}\ {\bar E}[\cos^2(\theta_0)\sin^{2d-2}(\theta_0)]. \\
\end{array}
$$

To prove item b), it is sufficient to check the result for $j=1$ and show that, due to the symmetry of the probability distribution, all the expected values are equal. Using again results from the Appendix we obtain:
$$
\begin{array}{ccl}
E[x_1^2] & = & E[\sin^2(\theta_0)\cos^2(\theta_1)] \\ \\
 & = & \displaystyle |{\cal S}_{2d-3}|\ \int_0^\pi f(\theta_0)\sin^{2d}(\theta_0)d\theta_0\ \int_0^\pi \cos^2(\theta_1)\sin^{2d-3}(\theta_1)d\theta_1 \\ \\
 & = & \displaystyle \dfrac{(2\pi)^{d-1}}{(2d-4)!!}\ {\bar E}[\sin^{2d}(\theta_0)]\ 2\dfrac{(2d-4)!!}{(2d-1)!!} = 2\dfrac{(2\pi)^{d-1}}{(2d-1)!!}\ {\bar E}[\sin^{2d}(\theta_0)]. \\
\end{array}
$$
Consider the intersection of the sphere ${\cal S}_{2d-1}$ with the hyperplane $x_0=\cos(\theta_0)$, for a fixed $0\leq \theta_0\leq \pi$. The result is a sphere ${\cal S}^{\prime}$ of dimension $2d-2$, radius $\sin(\theta_0)$ and center $(\cos(\theta_0),0,\dots,0)$. In this sphere the density function $f(\theta_0)$ is constant. Therefore the expected values $E_{{\cal S}^{\prime}}[x_j^2]$, $0<j<2d$, are all equal. This property is maintained when we integrate along the variable $\theta_0$. Then the expected values $E[x_j^2]$, $0<j<2d$, are also equal.

Items c) and d) follow from a) and b) and the following properties:
$$
\begin{array}{l}
E[|\alpha_0|^2] = E[x_0^2+x_1^2] = E[x_0^2] + E[x_1^2], \\ \\
E[|\alpha_k|^2] = E[x_{2k}^2+x_{2k+1}^2] = E[x_{2k}^2] + E[x_{2k+1}^2]. \\
\end{array}
$$
\end{proof}

Lemma~\ref{Lem:ExpectedCoordinates} allows us to calculate the expected value of $P_s$ for all $0\leq s< d^{\prime\prime}$.
\begin{cor}
\label{Cor:ExpectedProyections}
Given an isotropic random variable $\Psi$ with density function $f(\theta_0)$, the following statements hold true:
\begin{enumerate}
\item[a)] $\displaystyle E[P_0]=2\dfrac{(2\pi)^{d-1}}{(2d-3)!!}\left({\bar E}[\cos^2(\theta_0)\sin^{2d-2}(\theta_0)]+\dfrac{2d^\prime-1}{2d-1}{\bar E}[\sin^{2d}(\theta_0)]\right)$.
\item[b)] $\displaystyle E[P_0]=1-4\dfrac{(2\pi)^{d-1}}{(2d-1)!!}\ (d-d^\prime)\ {\bar E}[\sin^{2d}(\theta_0)]$.
\item[c)] $\displaystyle E[P_s]=4\dfrac{(2\pi)^{d-1}}{(2d-1)!!}\ d^\prime\ {\bar E}[\sin^{2d}(\theta_0)]$\quad for all\quad $0< s< d^{\prime\prime}$.
\end{enumerate}
\end{cor}
\begin{proof}
Items a) and c) are a direct consequence of Lemma~\ref{Lem:ExpectedCoordinates}. The proof of item b) is as follows:
$$
\begin{array}{ccl}
E[P_0] & = & 2\dfrac{(2\pi)^{d-1}}{(2d-3)!!}\left({\bar E}[\cos^2(\theta_0)\sin^{2d-2}(\theta_0)]+\dfrac{2d^\prime-1}{2d-1}{\bar E}[\sin^{2d}(\theta_0)]\right) \\ \\
 & = & 2\dfrac{(2\pi)^{d-1}}{(2d-3)!!}\left({\bar E}[\sin^{2d-2}(\theta_0)]-{\bar E}[\sin^{2d}(\theta_0)]+\dfrac{2d^\prime-1}{2d-1}{\bar E}[\sin^{2d}(\theta_0)]\right) \\ \\
 & = & 2\dfrac{(2\pi)^{d-1}}{(2d-3)!!}\left(\dfrac{(2d-3)!!}{2(2\pi)^{d-1}}-\dfrac{(2d-1)-(2d^\prime-1)}{2d-1}{\bar E}[\sin^{2d}(\theta_0)]\right) \\ \\
 & = & 1- 4\dfrac{(2\pi)^{d-1}}{(2d-1)!!}\ (d-d^\prime)\ {\bar E}[\sin^{2d}(\theta_0)]. \\
\end{array}
$$
To evaluate ${\bar E}[\sin^{2d-2}(\theta_0)]$ we have used Formula~(\ref{For:DensityFunction}).
\end{proof}

Given the state $\Psi$, to calculate the variance of the corrected state ${\tilde \Phi}$ we introduce the corrected states ${\tilde \Phi}_s$, $0\leq s< d^{\prime\prime}$, that are obtained when the measured syndrome is $s$, and we prove that all corrected states (${\tilde \Phi}$ and ${\tilde \Phi}_s$, $0\leq s< d^{\prime\prime}$) are isotropic. Note that $\Psi$, ${\tilde \Phi}$ and ${\tilde \Phi}_s$ ($0\leq s< d^{\prime\prime}$) are random variables, the first defined on the sphere ${\cal S}_{2d-1}$ and the others on the sphere ${\cal S}_{2d^\prime-1}$ contained in the subspace $S_0$.

\begin{thm}
\label{Thm:Isotropy}
Given an isotropic random variable $\Psi$, the random variables ${\tilde \Phi}$ and ${\tilde \Phi}_s$, $0\leq s< d^{\prime\prime}$, are isotropic. Furthermore, all of ${\tilde \Phi}_s$, $0< s< d^{\prime\prime}$, have a uniform distribution.
\end{thm}
\begin{proof}
To prove the result for $0< s< d^{\prime\prime}$ we resort to an argument similar to that used in Lemma~\ref{Lem:ExpectedCoordinates}. Consider the intersection of the sphere ${\cal S}_{2d-1}$ with the hyperplane $x_0=\cos(\theta_0)$, for a fixed $0\leq \theta_0\leq \pi$. The result is a sphere ${\cal S}^{\prime}$ of dimension $2d-2$, radius $\sin(\theta_0)$ and center $(\cos(\theta_0),0,\dots,0)$. In the sphere ${\cal S}^{\prime}$ the density function $f(\theta_0)$ is constant.

For any value of $0< s< d^{\prime\prime}$, we consider the sphere ${\cal S}^{\prime\prime}$ of points belonging to ${\cal S}^{\prime}$ whose projection belongs to the sphere of dimension $2d^\prime-1$ and radius $0\leq r\leq \sin(\theta_0)$ contained in the subspace $S_s$. Note that ${\cal S}^{\prime\prime}$ is like a parallel of ${\cal S}^{\prime}$ with respect to the equatorial subspace $S_s$. In the sphere ${\cal S}^{\prime\prime}$ the probability that the syndrome is $s$ is $r^2$ and hence it is constant. Then the projection of the points of ${\cal S}^{\prime\prime}$ on $S_s$ generates a uniform distribution in this subspace and, when applying the operator $E_s^{-1}$, a uniform distribution is obtained in $S_0$. This uniform distribution is maintained when we first integrate along the variable $r$ and then along the variable $\theta_0$. Finally, we conclude that the density function of ${\tilde \Phi}_s$ is constant and therefore this random variable is also constant. Therefore ${\tilde \Phi}_s$ has a uniform distribution for all values of $0< s< d^{\prime\prime}$.

The proof for $s=0$ is analogous. We consider the sphere ${\cal S}^{\prime}$ of points belonging to ${\cal S}_{2d-1}$ whose projection belongs to the sphere of dimension $2d^\prime-1$ and radius $0\leq r\leq 1$ contained in the subspace $S_0$. Note that ${\cal S}^{\prime}$ is like a parallel of ${\cal S}_{2d-1}$ with respect to the equatorial subspace $S_0$. In the sphere ${\cal S}^{\prime}$ the probability that the syndrome is $s=0$ is $r^2$ and hence it is constant. However, the density function $f(\theta_0)$ is not constant on ${\cal S}^{\prime}$. Now consider the sphere ${\cal S}^{\prime\prime}$ resulting from the intersection of ${\cal S}^{\prime}$ with the hyperplane $x_0=\cos(\theta_0)$, $|\cos(\theta_0)|\leq r$. Now the density function $f(\theta_0)$ is constant on ${\cal S}^{\prime\prime}$. Then the projection of the points of ${\cal S}^{\prime\prime}$ on $S_0$ generates an isotropic distribution in this subspace. This isotropic distribution is maintained when we first integrate along the variable $r$ and then along the variable $\theta_0$. Finally, we conclude that the density function of ${\tilde \Phi}_0$ is isotropic and therefore this random variable is also isotropic.

The isotropy of ${\tilde \Phi}$ is deduced from that of ${\tilde \Phi}_s$, $0\leq s< d^{\prime\prime}$, and from the fact that the density function of ${\tilde \Phi}$ is the sum of the density functions of ${\tilde \Phi}_s$ for all $0\leq s< d^{\prime\prime}$. Note that the density functions of ${\tilde \Phi}_s$, $0\leq s< d^{\prime\prime}$, have total weight less than $1$ (their integral is less than $1$), but when added together they have total weight equal to $1$.
\end{proof}

Given a specific $\Psi$, the variance of ${\tilde \Phi}$, $V({\tilde \Phi})=E[\|\Phi-{\tilde \Phi}\|^2]$, is calculated as follows, bearing in mind the results of Theorem~\ref{Thm:Isotropy}:
\begin{equation}
\label{FinalVariance}
\begin{array}{ccl}
V({\tilde \Phi}) & = & E[P_0 \|\Phi-E_0^{-1}\Pi_0\Psi\|^2] + (d^{\prime\prime}-1)E[P_1 \|\Phi-E_1^{-1}\Pi_1\Psi\|^2] \\ \\
 & = & E[P_0 \|\Phi-\Pi_0\Psi\|^2] + (d^{\prime\prime}-1)E[P_1 \|E_1\Phi-\Pi_1\Psi\|^2].
\end{array}
\end{equation}
where $\Pi_0$ and $\Pi_1$ are the projectors on the subspaces $S_0$ and $S_1$ respectively. In the last equality we have used that $E_0=I$ and that $E_1$ is a unitary transformation and, therefore, preserves the norm. We first calculate the value of the expressions $P_0 \|\Phi-\Pi_0\Psi\|^2$ and $P_1 \|E_1\Phi-\Pi_1\Psi\|^2$.


\begin{lem}
\label{Lem:PuntualError}
The following equalities hold true:

$\quad\displaystyle
\begin{array}{ccl}
a)\ P_0\|\Phi-\Pi_0\Psi\|^2 & = & 2-2\sin^2(\theta_0)\cdots\sin^2(\theta_{2d'-1})\,  \\ \\
 & & -2\cos(\theta_0)\sqrt{1-\sin^2(\theta_0)\cdots\sin^2(\theta_{2d'-1})}\ , \\ \\
\end{array}
$

$\quad\displaystyle
\begin{array}{l}
b)\ P_1\|E_1\Phi-\Pi_1\Psi\|^2 = 
2\sin^2(\theta_0)\cdots\sin^2(\theta_{2d'-1})\  \\ \\
\qquad -2\sin^2(\theta_0)\cdots\sin^2(\theta_{2d'-1})\sin^2(\theta_{2d'})\cdots\sin^2(\theta_{4d'-1})\  \\ \\
\qquad -2\sin^2(\theta_0)\cdots\sin^2(\theta_{2d'-1})\cos(\theta_{2d'})\sqrt{1-\sin^2(\theta_{2d'})\cdots\sin^2(\theta_{4d'-1})}\ . \\
\end{array}
$
\end{lem}
\begin{proof}
Taking into account that $S_0=L(|0\rangle,\dots,|d^\prime-1\rangle)$, the probability that the state $\Psi$, described by Formula (\ref{For:BinomialForm}), is projected onto $S_0$ is:
$$
\begin{array}{lcl}
P_0 & = & \cos^2(\theta_0)+\sin^2(\theta_0)\cos^2(\theta_1)+\cdots +\sin^2(\theta_0)\cdots \sin^2(\theta_{2d^\prime-2})\cos^2(\theta_{2d^\prime-1}) \\ \\
    & = & 1-\sin^2(\theta_0)\cdots \sin^2(\theta_{2d^\prime-2}) \sin^2(\theta_{2d^\prime-1}). \\
\end{array}
$$
The last equality is obtained by substituting $\cos^2(\theta_{2d^\prime-1})$ for $1-\sin^2(\theta_{2d^\prime-1})$ and observing that the positive terms add up to $1$, because they correspond to the parametrization of a point in a sphere of dimension $2d^\prime-1$.

On the other hand, taking into account that $\Phi=|0\rangle$,
$$
\begin{array}{lcl}
\displaystyle P_0\|\Phi-\Pi_0\Psi\|^2 & = & \displaystyle  P_0\left\||0\rangle-\dfrac{1}{\sqrt{P_0}}\sum_{k=0}^{d^\prime-1}\alpha_k |k\rangle\right\|^2 = \left\|\sqrt{P_0}|0\rangle-\sum_{k=0}^{d^\prime-1}\alpha_k |k\rangle\right\|^2 \\ \\
    & = & \displaystyle \left|\sqrt{P_0}-\alpha_0\right|^2+\sum_{k=1}^{d^\prime-1}|\alpha_k|^2 = 2P_0-2cos(\theta_0)\sqrt{P_0}. \\
\end{array}
$$

Substituting the value of $P_0$ obtained previously in the last expression, the equality a) is proved.

An analogous development allows us to prove the equality b). In this way the proof is concluded.
\end{proof}

Next, we compute the expected values of the expressions obtained in Lemma~\ref{Lem:PuntualError}.

\begin{lem}
\label{Lem:ExpectedValues}
The following equalities hold true:
$$
\begin{array}{lcl}
a)\ E_1&=&E[\sin^2(\theta_0)\cdots\sin^2(\theta_{2d'-1})] = 4\,\dfrac{(2\pi)^{d-1}}{(2d-1)!!}\,(d-d^\prime)\,{\bar E}[\sin^{2d}(\theta_0)]. \\ \\
b)\ E_2&=&E[\sin^2(\theta_0)\cdots\sin^2(\theta_{4d'-1})] = 4\,\dfrac{(2\pi)^{d-1}}{(2d-1)!!}\,(d-2d^\prime)\,{\bar E}[\sin^{2d}(\theta_0)]. \\ \\
c)\ E_3&=&E\left[\sin^2(\theta_0)\cdots\cos(\theta_{2d'})\sqrt{1-\sin^2(\theta_{2d'})\cdots\sin^2(\theta_{4d'-1})}\right] = 0. \\ \\
d)\ E_4&=&E[\cos(\theta_0)\sqrt{1-\sin^2(\theta_0)\cdots\sin^2(\theta_{2d'-1})}]  \\ \\
&=& 2\,\dfrac{(2\pi)^{d-1}}{(2d-3)!!}\,{\bar E}[\cos(\theta_0)\sin^{2d-2}(\theta_0)] - 2\,\dfrac{(2\pi)^{d-1}}{(2d-2d^\prime-2)!!} \\ \\
& & \cdot  \displaystyle \sum_{k=1}^{\infty} \dfrac{(2k-3)!!(2d-2d^\prime+2k-2)!!}{(2k)!!(2d+2k-3)!!} \,{\bar E}[\cos(\theta_0)\sin^{2d+2k-2}(\theta_0)].\\
\end{array}
$$
\end{lem}
\begin{proof}
The equalities a) and b) can be proved by expanding the integrals in the expected value and using the results of the Appendix.
$$
\begin{array}{lcl}
E_1 & = & \displaystyle |{\cal S}_{2d-2d^\prime-1}|\,{\bar E}[\sin^{2d}(\theta_0)] \\ \\
 & & \cdot \displaystyle \int_0^{\pi} \sin^{2d-1}(\theta_1)d\theta_1 \cdots \int_0^{\pi} \sin^{2d-2d^\prime+1}(\theta_{2d^\prime-1})d\theta_{2d^\prime-1}, \\ \\
E_2 & = & \displaystyle |{\cal S}_{2d-4d^\prime-1}|\,{\bar E}[\sin^{2d}(\theta_0)] \\ \\
 & & \cdot \displaystyle \int_0^{\pi} \sin^{2d-1}(\theta_1)d\theta_1 \cdots \int_0^{\pi} \sin^{2d-4d^\prime+1}(\theta_{4d^\prime-1})d\theta_{4d^\prime-1}. \\
\end{array}
$$

In c), the third expected value equals $0$ because the function is odd (with respect to the midpoint of the interval $[0,\pi]$) for the variable $\theta_{2d'}$.

In d), the expected value is expressed as a function of an integral over the hemisphere ${\bar{\cal S}}_{2d^\prime}$ (the last polar angle $\theta_{2d^\prime-1}$ takes values on $[0,\pi]$ instead of $[0,2\pi]$). Note that in the hemisphere ${\bar{\cal S}}_{2d^\prime}$ the last coordinate ($\sin(\theta_0)\cdots \sin(\theta_{2d^\prime-1})$) is always positive.
$$
\begin{array}{lcl}
E_4 & = & \displaystyle |{\cal S}_{2d-2d^\prime-1}| \int_{{\bar{\cal S}}_{2d^\prime}} f(\theta_0) \cos(\theta_0) \\ \\
 & & \cdot \displaystyle \sqrt{1-(\sin(\theta_0)\cdots \sin(\theta_{2d^\prime-1}))^2} (\sin(\theta_0)\cdots \sin(\theta_{2d^\prime-1}))^{2d-2d^\prime-1} d_{{\bar{\cal S}}_{2d^\prime}}. \\
\end{array}
$$

The integrand above depends exclusively on the first and last coordinates, $\cos(\theta_0)$ and $\sin(\theta_0)\cdots \sin(\theta_{2d^\prime-1})$ respectively. We perform a change of reference system in the space $R^{2d^\prime+1}$ in which the hemisphere ${\bar{\cal S}}_{2d^\prime}$ is contained, cyclically rotating the second, penultimate and last coordinate axes. In this reference system, the function that we want to integrate depends exclusively on the first and second coordinates, $\cos(\theta_0)$ and $\sin(\theta_0)\cos(\theta_1)$ respectively. Furthermore, the second polar angle only takes values in the interval $[0,\pi/2]$ since now the second coordinate must always be positive. With this change the previous equality is transformed as follows:
$$
\begin{array}{lcl}
E_4 & = & \displaystyle |{\cal S}_{2d-2d^\prime-1}|\ |{\cal S}_{2d^\prime-2}| \\ \\
 & & \cdot \displaystyle \int_0^{\pi} I(\theta_0)\, f(\theta_0) \cos(\theta_0) \sin^{2d-2d^\prime-1}(\theta_0) \sin^{2d^\prime-1}(\theta_0) d\theta_0 \qquad\text{where} \\ \\
I(\theta_0) & = & \displaystyle \int_0^{\frac{\pi}{2}} \sqrt{1-\sin^2(\theta_0)\cos^2(\theta_1)} \cos^{2d-2d^\prime-1}(\theta_1) \sin^{2d^\prime-2}(\theta_1) d\theta_1. \\
\end{array}
$$

The proof is concluded in a simple way using the following development together with the results of the Appendix:
$$
\sqrt{1-x^2} = 1-\sum_{k=1}^{\infty} \dfrac{(2k-3)!!}{(2k)!!} x^{2k}.
$$
\end{proof}

Finally we are able to calculate the variances of the corrected state ${\tilde \Phi}$.

\begin{thm}
\label{Thm:FinalVariance}
The variance of the random variable ${\tilde \Phi}$ satisfies:
$$
\begin{array}{lcl}
V({\tilde \Phi}) & = & \displaystyle 2-4 \dfrac{(2\pi)^{d-1}}{(2d-3)!!} \,{\bar E}[\cos(\theta_0)\sin^{2d-2}(\theta_0)] + 4 \dfrac{(2\pi)^{d-1}}{(2d-2d^\prime-2)!!} \\ \\
 & & \cdot \displaystyle \sum_{k=1}^{\infty} \dfrac{(2k-3)!!(2d-2d^\prime+2k-2)!!}{(2k)!!(2d+2k-3)!!} \,{\bar E}[\cos(\theta_0)\sin^{2d+2k-2}(\theta_0)]. \\ \\
\end{array}
$$
\end{thm}
\begin{proof}
Just use the results of Lemmas~\ref{Lem:PuntualError} and~\ref{Lem:ExpectedValues} in Formula~(\ref{FinalVariance}).
\end{proof}

The results of Theorems~\ref{Thm:Variance} and~\ref{Thm:FinalVariance} allow us to show, for very general cases, that the variance of the corrected state, $V({\tilde \Phi})$, is greater than the variance of the disturbed state, $V(\Psi)$. Therefore, the conclusion is that no quantum error correcting code fix isotropic errors.

\begin{figure}[th]
\label{Fig:Cos_Sin}
\begin{center}
        \includegraphics[scale=1]{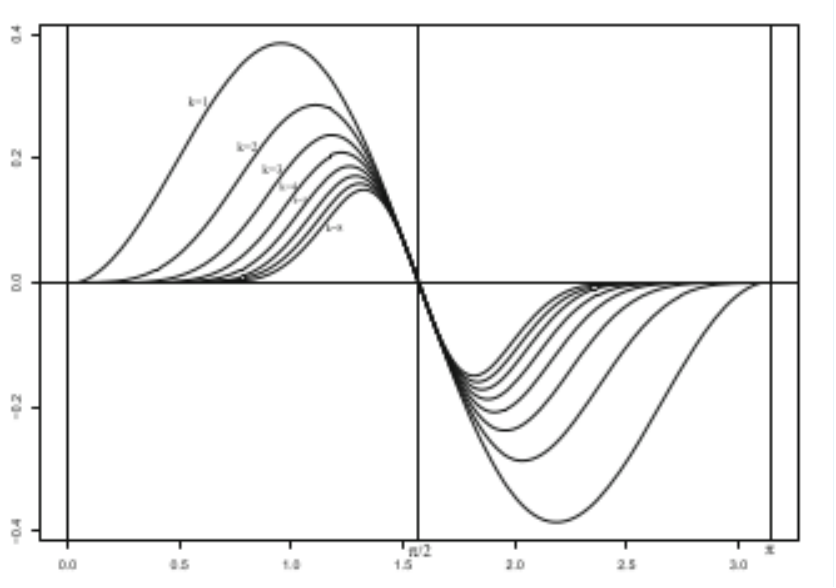}
        \caption{\centerline{Representations of function $\cos(\theta_0)\sin^{2k}(\theta_0)$.}}
\end{center}
\end{figure}

\begin{thm}
\label{Thm:FinalResult}
Let $\Psi$ be an isotropic random variable with a density function $f(\theta_0)$. Then $V({\tilde \Phi}) - V(\Psi)\geq 0$ if $f(\theta_0)$ satisfies any of the following properties:
\begin{enumerate}
\item[a)] $f(\theta_0)$ is a non-increasing function.
\item[b)] $f(\theta_0)=0$ for all $\theta_0\geq \pi/2$.
\item[c)] $f(\pi/2-\theta)\geq f(\pi/2+\theta)$ for all $0\leq \theta\leq \pi/2$.
\end{enumerate}
\end{thm}
\begin{proof}
Substituting the result of Theorem~\ref{Thm:Variance} in the formula given in the Theorem~\ref{Thm:FinalVariance}, we obtain:
$$
\begin{array}{lcl}
V({\tilde \Phi}) - V(\Psi) & = & \displaystyle 4 \dfrac{(2\pi)^{d-1}}{(2d-2d^\prime-2)!!} \\ \\
 & & \cdot \displaystyle \sum_{k=1}^{\infty} \dfrac{(2k-3)!!(2d-2d^\prime+2k-2)!!}{(2k)!!(2d+2k-3)!!} \,{\bar E}[\cos(\theta_0)\sin^{2d+2k-2}(\theta_0)]. \\ \\
\end{array}
$$

The above variance difference is positive if any of the properties stated in the theorem is satisfied (See Figure~\ref{Fig:Cos_Sin}), since any of them is enough to guarantee that:
$$
{\bar E}[\cos(\theta_0)\sin^{2d+2k-2}(\theta_0)] \geq 0\quad\text{for all}\ k\geq 1.
$$
\end{proof}

\subsection{Example}

The normal distribution introduced in Definition~\ref{Sub:Example} satisfies the following properties:

$$
\begin{array}{l}
E[x_0^2] = \dfrac{1+(2d-1)\sigma^2}{2d}, \\ \\
E[x_j^2] =  \dfrac{1-\sigma^2}{2d} \quad\text{for all}\ 1< j< 2d, \\ \\
E[|\alpha_0|^2] = \dfrac{1+(d-1)\sigma^2}{d}, \\ \\
E[|\alpha_k|^2] = \dfrac{1-\sigma^2}{d} \quad\text{for all}\ 1< k< d, \\ \\
E[P_0] = 1 - \dfrac{d^{\prime\prime}-1}{d^{\prime\prime}}(1-\sigma^2), \\ \\
E[P_s] = \dfrac{(1-\sigma^2)}{d^{\prime\prime}} \quad\text{for all}\ 1< s< d^{\prime\prime}. \\ \\
\end{array}
$$

They are obtained from Lemma~\ref{Lem:ExpectedCoordinates} and Corollary~\ref{Cor:ExpectedProyections}, using the results of the Appendix.

If the isotropic random variable $\Psi$ has a normal distribution with density function $f_n(\sigma,\theta_0)$ ($0\leq \sigma< 1$) then, since it satisfies the first and third of the conditions of Theorem~\ref{Thm:FinalResult}, it holds:
$$
V({\tilde \Phi}) - V(\Psi) > 0.
$$

\section{Conclusions}

Quantum isotropic errors are simple enough to allow a complete study. However, they do not occur as a consequence of errors in quantum gates and quantum measurements. An important feature of quantum errors introduced in~\cite{LP}, in addition to variance, is the shape of their density functions, particularly the dimension of their supports (set of points in the domain where the density function is greater than zero). For example, the support of a density function that represents the error in a qubit has dimension $4$, far from the support dimension of an isotropic error that is equal to $d=2^{n+1}-1$. On the other hand, it is obvious that the dimension of the support of the sum of two independent errors in less than or equal to the sum of the dimensions of the respective supports. This property may cause the dimension of the error support to grow. However, this growth has limits. Thereby, the local quantum computing error model imposes a limitation on the growth of the dimension of the error support. For example, the support of independent errors in each qubit will have a maximum dimension of $4n$.

Despite the difference in the dimension of the support between isotropic errors and qubit independent errors, both types of errors have a global behavior. The proof of Theorem~\ref{Thm:Isotropy} provides a rather surprising intermediate result: the probability distribution of the final state $\tilde\Phi_s$ when an $s$ syndrome is detected, $0<s<d^{\prime\prime}$, is uniform. This means that if an error is detected in the code correcting circuit, all the logical $m-$qubit information has already been lost in computing. This result is not surprising for isotropic errors. What is striking is that it also occurs for qubit independent errors, according to the ongoing investigations that we are carrying out. This fact indicates that errors with a relatively small support dimension ($4n$ versus $2^{n+1}-1$) have an overall behavior.

In view of these results, it is important to determine which types of quantum computing errors can be fixed by quantum error correcting codes and which cannot, that is, which errors are controllable and which are not. In this work we have proven that isotropic errors are in the class of uncontrollable errors.

\section{Appendix}

The values of the integrals that have been used throughout the article are included in this appendix:

$$
\begin{array}{l}
\displaystyle \int_{0}^{\pi}\sin^{k}(\theta)d\theta=
\left\{\begin{array}{l}
\displaystyle 2\,\frac{(k-1)!!}{k!!}\quad\mbox{}\quad k=1,\,3,\,5,\,\,\dots \\ \\
\displaystyle \pi\,\frac{(k-1)!!}{k!!}\quad\mbox{}\quad k=2,\,4,\,6,\,\,\dots
\end{array}\right., \\ \\
\displaystyle \int_{0}^{\frac{\pi}{2}}\cos^a(\theta)\,\sin^b(\theta)\,d\theta=
\left\{\begin{array}{l}
\displaystyle\frac{\pi}{2}\frac{(a-1)!!\,(b-1)!!}{(a+b)!!}\quad\mbox{if $a$ and $b$ are even} \\ \\
\displaystyle\frac{(a-1)!!\,(b-1)!!}{(a+b)!!}\quad\mbox{in another case}
\end{array}\right., \\ \\
\displaystyle \int_{0}^{\pi}\frac{\sin^{2d-2}(\theta_0)}{(1+\sigma^2-2\sigma\cos(\theta_0))^d}d\theta_0=
\frac{(2d-3)!!}{(2d-2)!!}\frac{\pi}{(1-\sigma^2)}\quad\mbox{}\quad d=1,\,2,\,3,\,\,\dots\, , \\ \\
\displaystyle \int_{0}^{\pi}\frac{\cos(\theta_0)\sin^{2d-2}(\theta_0)}{(1+\sigma^2-2\sigma\cos(\theta_0))^d}d\theta_0=
\frac{(2d-3)!!}{(2d-2)!!}\frac{\sigma}{(1-\sigma^2)}\pi\quad\mbox{}\quad d=1,\,2,\,3,\,\,\dots\, , \\ \\
\displaystyle \int_{0}^{\pi}\frac{\sin^{2d}(\theta_0)}{(1+\sigma^2-2\sigma\cos(\theta_0))^d}d\theta_0= \frac{(2d-1)!!}{(2d)!!}\pi\quad\mbox{}\quad d=0,\,1,\,2,\,\,\dots\, , \\ \\
\displaystyle \int_{0}^{\pi}\frac{\cos^2(\theta_0)\sin^{2d-2}(\theta_0)}{(1+\sigma^2-2\sigma\cos(\theta_0))^d}d\theta_0 = \pi\,\frac{(2d-3)!!}{(2d)!!}\,\frac{1+(2d-1)\sigma^2}{1-\sigma^2} \ \mbox{}\ d=1,\,2,\,3,\,\,\dots\, . \\ \\
\end{array}
$$

Starting from the first integral, the surface of a unit sphere of arbitrary even ($2d$) or odd ($2d-1$) dimension can be calculated:

$$
\begin{array}{l}
\begin{array}{ccl}
|{\cal S}_{2d}| & = & \displaystyle \int_0^{\pi}\cdots\int_0^{\pi}\int_0^{2\pi}\sin^{2d-1}(\theta_0)\,\cdots\,\sin^{1}(\theta_{2d-2})\ d\theta_0\,\cdots\,d\theta_{2d-2}d\theta_{2d-1} \\ \\
 & = & \displaystyle 2\frac{(2d-2)!!}{(2d-1)!!}\ \frac{(2d-3)!!}{(2d-2)!!}\pi\ 2\frac{(2d-4)!!}{(2d-3)!!}\ \cdots\ \frac{(2-1)!!}{2!!}\pi\ 2\frac{(1-1)!!}{1!!}\ 2\pi \\ \\
 & = & \displaystyle \frac{2(2\pi)^d}{(2d-1)!!}, \\
\end{array} \\ \\
\begin{array}{ccl}
|{\cal S}_{2d-1}| & = & \displaystyle \int_0^{\pi}\cdots\int_0^{\pi}\int_0^{2\pi}\sin^{2d-2}(\theta_0)\,\cdots\,\sin^{1}(\theta_{2d-3})\ d\theta_0\,\cdots\,d\theta_{2d-3}d\theta_{2d-2} \\ \\
 & = & \displaystyle \frac{(2d-3)!!}{(2d-2)!!}\pi\ 2\frac{(2d-4)!!}{(2d-3)!!}\ \cdots\ \frac{(2-1)!!}{2!!}\pi\ 2\frac{(1-1)!!}{1!!}\ 2\pi \\ \\
 & = & \displaystyle \frac{(2\pi)^d}{(2d-2)!!}. \\
\end{array} \\
\end{array}
$$

\end{document}